1    **Suppression of the water ice and snow albedo feedback on planets orbiting**

2    **red dwarf stars and the subsequent widening of the habitable zone**

4    Manoj M. Joshi* and Robert M. Haberle**

5    *NCAS Climate, University of Reading

6    **Space Science and Astrobiology Division, NASA Ames Research Centre

8    Corresponding author address:

9    Manoj Joshi

10   NCAS Climate

11   Department of Meteorology

12   University of Reading

13   PO Box 243

14   Reading RG6 6BB

15   United Kingdom

16   Email: m.m.joshi@reading.ac.uk

17   Phone: +44 (0)118 378 7991

18   Fax: +44 (0)118 378 8905

20   **Running title:** snow/ice albedo and M-dwarf habitability

22    **Abstract**

23    M-stars comprise 80% of main-sequence stars, and so their planetary systems

24    provide the best chance for finding habitable planets, that is, those with surface

25    liquid water. We have modeled the broadband albedo or reflectivity of water ice

26    and snow for simulated planetary surfaces orbiting two observed red dwarf stars

27    (or M-stars), using spectrally resolved data of the Earth's cryosphere. The gradual

28    reduction of the albedos of snow and ice at wavelengths greater than 1 µm,

29    combined with M-stars emitting a significant fraction of their radiation at these

30    same longer wavelengths, mean that the albedos of ice and snow on planets

31    orbiting M-stars are much lower than their values on Earth. Our results imply that

32    the ice/snow albedo climate feedback is significantly weaker for planets orbiting

33    M-stars than for planets orbiting G-type stars such as the Sun. In addition, planets

34    with significant ice and snow cover will have significantly higher surface

35    temperatures for a given stellar flux if the spectral variation of cryospheric albedo

36    is considered, which in turn implies that the outer edge of the habitable zone

37    around M-stars may be 10-30% further away from the parent star than previously

38    thought.

39

**Introduction**

The habitability of extrasolar planets is a question that has stimulated many observational and theoretical studies over the past few decades, with recent actual observations of planets increasing this interest to new levels. The habitability of planets orbiting small red dwarfs, or M-stars, is a question that demands special attention because M-stars comprise 80% of main-sequence stars, and so their planetary systems provide the best chance for finding habitable planets, that is, those with surface liquid water.

The so-called habitable zone, or the loci of orbits where liquid water is stable around a range of different stars, has been investigated by several authors (e.g., Huang 1959, Dole 1964, Kasting et al 1993). They pointed out that tidal locking-- the phenomenon whereby one side always faces the parent star--could present a barrier to the habitability of planets orbiting M-stars. However, this issue has been addressed in models ranging in complexity from simple energy-balance systems (Haberle et al. 1996) to complex three-dimensional global circulation models (Joshi et al 1997). More recently, climate models that include radiative processes such as clouds and water vapor have been employed to examine the problem both in general (Joshi 2003) and for specific planets orbiting specific stars such as Gliese 581 (Pierrehumbert 2011). The issues of M-stars emitting flares (sudden releases of energy such as X-rays or ultraviolet radiation) and the chemical

62  composition of planets orbiting M-stars has also been addressed recently (Scalo et

63  al. 2007, Segura et al. 2010). Elsewhere, the habitability of planets orbiting M-

64  stars has been examined in a comprehensive review paper (Tarter et al 2007).

65

66  In the present work, we address a property of water ice and snow that has

67  significant implications for the M-star habitability question, and that is the

68  dependence of ice and snow albedo on wavelength. The spectral dependence of

69  snow and ice albedo has been observed and reported upon in previous work (e.g.,

70  Ebert and Curry 1993, Brandt et al 2005, Hudson et al 2006), and some GCMs do

71  incorporate such effects by splitting up the shortwave radiation into two

72  components separated at 0.7-0.9 μm (e.g., Dickinson et al 1986). However, this

73  dependence is small for wavelengths shorter than 1 μm, where the vast majority

74  of the Sun's energy is emitted, so the effect on terrestrial climate is small.

75

76  M-stars are much smaller and cooler than G-stars, such as the Sun, and as a

77  consequence emit a far greater fraction of their radiation at wavelengths longer

78  than 1 μm. Figure 1 (top panel) shows black body spectra for the Sun (shown in

79  black), for an idealized black body emitting at 3300K (red), and for two actual M-

80  stars (purple and orange). The two M-stars emit a large fraction of radiation at

81  wavelengths longer than 1 μm where the snow and ice albedos are smaller than in

82  the visible. This implies that, because of the spectral dependence of ice and snow

83 albedo on wavelength, a disproportionate amount of the long wave radiation
84 emitted by such M-stars will be absorbed rather than reflected from an icy or
85 snowy surface, which would thus lower the average albedo of such a surface.

86

87 Spectra of M-stars have been obtained from the "NextGen" stellar atmospheres
88 grid of Hauschildt et al. (1999). The spectra are significantly non-Planckian, as
89 shown by the purple and orange curves in Figure 1 (top panel): Gliese 436 and GJ
90 1214 emit significantly more radiation in the 3-10 μm region than would be
91 expected from black bodies having the same temperatures as these stars. In the
92 next section, we quantify how the spectral variation of ice and snow albedo
93 affects the mean bolometric values.

94

95 **Results**

96 We examined the effect described above by asking the question, "What would be
97 the average albedo of snow and ice receiving radiation having the spectral
98 characteristics of the red, purple, and orange curves in Figure 1 (top panel)?" We
99 used data for water ice from Figure 1 of Brandt et al (2005), and snow data from
100 Figure 5 of Hudson et al (2006). The data as used in our model is shown in Figure
101 1 (bottom panel).

102

103　Using the albedo information above, and the spectral emission functions shown in

104　Figure 1 (top panel), we calculated broadband albedos as shown in Figure 2. The

105　values for snow and ice for a planetary surface orbiting the Sun are 0.8 and 0.5,

106　respectively, which are broadly consistent with the values used in climate models.

107　Fresh snow and ice albedos on a planet receiving black body radiation from an

108　object at 3300K are 0.6 and 0.3, respectively, which are significant reductions

109　from the "solar" values.

110

111　When broadband albedos are calculated for fresh snow and ice for surfaces

112　receiving radiation from the two stars Gliese 436 and GJ 1214, the snow albedos

113　are lowered significantly to 0.47 and 0.43, respectively; the ice albedos are also

114　lowered significantly more to 0.24 and 0.23, respectively.

115

116　**Discussion- Climate Feedbacks**

117　The lowering of average water ice and snow albedos has implications for the

118　climates of planets orbiting M-stars. For instance, consider a planet similar to the

119　example of Pierrehumbert et al (2011), but which orbits Gliese 436, and has half

120　the incident radiation incident on open ocean with an albedo of 0.1 and half

121　incident on ice. The average surface albedo of the planet will be changed from

122　(0.1+0.50)/2.0 or 0.30, to (0.1+0.24)/2.0 or 0.17, if the spectral dependence of ice

123　albedo is considered. As another example, if the same planet were to have 50%

124  land cover that had a bare-ground albedo of 0.2, and the same distribution of ice

125  and snow as in the previous example, the average surface albedo of the planet

126  would be changed from (0.10+0.50+0.2+0.8)/4.0 or 0.40 to

127  (0.10+0.24+0.2+0.47)/4.0 or 0.25.

128

129  The reduction in albedo weakens the strength of the snow/ice albedo feedback,

130  which is a mechanism whereby a perturbation to snow or ice coverage associated

131  with a climate forcing results in a change in the amount of stellar radiation

132  absorbed by the surface due to the large difference between the albedos of snow

133  or ice, and the albedos of the underlying ground or ocean. The snow/ice albedo

134  feedback is a positive feedback, because any change to snow or ice cover

135  amplifies the climatic forcing that caused the change in the snow/ice cover in the

136  first place. For instance, a forcing that warms a planet will cause some snow or

137  ice to melt, which will reduce planetary albedo and result in additional warming

138  (and vice versa).

139

140  While the exact change in the snow/ice albedo feedback will depend on the

141  specific nature of the planetary climate system being studied, an approximate

142  number can be calculated by using the common terminology for climate

143  feedbacks in Earth's atmosphere (Colman 2003, Gregory et al 2009):

144  $$dT = dF \, / \, (B + W + C + I) \qquad (1)$$

145      In this terminology, the temperature response $dT$ to a radiative forcing $dF$ is given

146      by the black body response $B$ (3.3 Wm$^{-2}$K$^{-1}$) modulated by a number of terms,

147      which are negative if they are positive feedbacks. These terms are the water vapor

148      feedback $W$ ($\approx$ -1.5 Wm$^{-2}$K$^{-1}$), the cloud feedback $C$ (-0.75 to +0.75 Wm$^{-2}$K$^{-1}$),

149      and the ice albedo feedback $I$ ($\approx$ -0.3 Wm$^{-2}$K$^{-1}$). If the changes in albedo above

150      mean that $I$ is small, then the clear-sky feedback parameter $(B + W + I)$ changes

151      from 1.5 Wm$^{-2}$K$^{-1}$ to $\approx$ 1.8 Wm$^{-2}$K$^{-1}$, which thus reduces the total climate response

152      to a given perturbation.

153

154      While the reduction of ice albedo has a small effect on the climate sensitivity of a

155      planetary climate that has similar characteristics to present-day Earth, its effect

156      may be expected to be more profound on planets that are largely ice-covered, in a

157      similar manner to the so-called "snowball Earth" period (e.g., Kirschvink 1992).

158      Previous studies suggest that the snowball Earth scenario might be indicative of a

159      bistable state (Budyko 1969), where a planet on the cusp of a runaway glaciation

160      can be pushed into an ice-covered state by a small climatic perturbation and a

161      strong ice/snow albedo feedback. The lowering of snow and ice albedo to near

162      bare-ground or open-ocean values therefore makes runaway glaciation and

163      "snowball-planet" episodes much less likely.

164

165

166 **Discussion- Habitable Zone**

167 We estimate the effect of albedo variation on the width of the habitable zone by

168 calculating downward top-of-atmosphere (TOA) stellar flux $F$ vs albedo for a

169 planetary surface temperature of 200K. This is the temperature at which a bar of

170 $CO_2$ condenses, and it is often considered a proxy to the outer edge of the

171 habitable zone (Kasting et al. 1993). We consider the black body relationship:

172 $$F = \varepsilon\sigma T^4/(1\text{-}a) \qquad\qquad (2)$$

173 where $F$ is the downward top-of-atmosphere stellar flux, $\sigma$ is the Stefan-

174 Boltzmann constant, $\varepsilon$ is the emissivity, $a$ is surface albedo, and $T$ is surface

175 temperature. We assume that at a surface temperature of 200K the amount of

176 water vapor in the atmosphere is negligible, so any deviation of $\varepsilon$ from unity is

177 due to $CO_2$. We assume that $\varepsilon = 0.8$, consistent with a bar of $CO_2$ having a grey

178 optical depth of approximately unity (Joshi et al 1997). We neglect clouds and

179 assume that Rayleigh scattering can be ignored to leading order as its effects on

180 planets receiving radiation from M-dwarfs is small (Kasting et al 1993).

181

182 Figure 3 (left panel) shows albedo vs downward TOA stellar flux for the above

183 simple model when T = 200K. Consideration of the spectral dependence of ice

184 albedo considerably changes the TOA flux at which surface temperature equals

185 200 K (our proxy for the habitable zone edge). Figure 3 (right hand panel) shows

186 the same result as Figure 3 (left panel), but it is expressed in terms of orbital

187    distance in AU away from the parent star, which enables a quantification of the

188    effect of the spectral variation in ice/snow albedo on habitable zone width. For

189    instance, considering the example of partially ice covered planets in the previous

190    section, decreasing albedo from 0.30 to 0.17 increases the orbital distance at

191    which $CO_2$ condensation happens from 1.8 to 1.98 AU, or 10%, while decreasing

192    albedo on planets covered by 50% snow and 50% ice from 0.65 to 0.33 increases

193    the same orbital distance from 1.3 to 1.75 AU, or over 30%.

194

195    The above calculations all assume two ice and snow types from the works of

196    Brandt et al (2005) and Hudson et al (2006), respectively. However, different ice

197    and snow types have different albedos. For instance, glacier ice albedo decreases

198    from $\approx 0.9$ at 0.5 μm to $\approx 0.1$ at 1 μm (Warren et al 2002). Pure water-ice formed

199    from fresh water, as occurs on fresh water lakes, can have a very low albedo that

200    is almost independent of wavelength (e.g., Bolsenga 1969). Such ice would

201    eliminate the presence of the ice-albedo feedback completely. Indeed, at very low

202    temperatures approaching 200 K, cubic crystalline water ice can form as opposed

203    to common hexagonal crystalline water ice, which might have implications for ice

204    albedo on the outer edge of the habitable zone. Only with observations of the

205    surfaces of exoplanets could investigators discern the type of ice or snow that

206    exists on them.

207

208    The above calculations assume an ocean-covered planet and no clouds. $CO_2$ and

209    water vapor have absorption bands between 1 and 10 μm, and decrease the

210    fraction of incoming radiation that reaches the surface of a planet orbiting an M-

211    star at these wavelengths. We approximately quantify such effects by repeating

212    our calculations, but multiplying the amplitude of the spectral distribution

213    function in Figure 1 (top panel) by half at wavelengths above 1.5 μm. The effects

214    on ice albedo are small, perhaps unsurprisingly, since Figure 2 indicates that the

215    albedo of ice drops to low values below 1 μm. The effect on snow albedo is more

216    significant and raises the weighted snow albedo in the Gliese 436 and GJ 1214

217    cases by 0.1.

218

219    In the future, more detailed calculations will have to be carried out by using 3D

220    models of planets orbiting specific stars, such as that of Pierrehumbert (2011).

221    Such models are able to take account of the above effects, as well as the variation

222    of stellar zenith angle, and the effect of different atmospheric gaseous absorbers.

223    Nevertheless, we believe the above result, which is that the outer edge of the

224    habitable zone moves outward when the spectral dependence of ice and snow

225    albedo is taken into account, is robust.

226

227    $CO_2$ clouds can potentially provide an effective scattering greenhouse effect (e.g.,

228    Forget and Pierrehumbert 1997) and could be present on a planet near the outer

229   edge of the habitable zone. The potential effects of such clouds overlying ice or
230   snow with spectrally varying albedos would depend on atmospheric and
231   microphysical parameters, and as such these effects are beyond the scope of this
232   paper. We suggest quantifying such impacts as another source for future research.

233

234   The effect considered here should not move the inner edge of the habitable zone,
235   usually considered as the locus of orbits where loss rates of water become
236   significant to dry a planet on geological timescales (Kasting et al 1993), away
237   from the parent M-star. This is because when a planet is at the inner edge of the
238   habitable zone, surface temperatures should be high enough to ensure that ice
239   cover is small. For a tidally locked planet, this implies that ice is confined to the
240   dark side that perpetually faces away from the parent star; such ice receives no
241   stellar radiation, which renders albedo effects unimportant.

242

243   It has been suggested that methane ($CH_4$) and nitrous oxide ($N_2O$) are more stable
244   in planetary atmospheres orbiting M-stars than they are on Earth due to the very
245   low amount of ultraviolet radiation emitted by such stars that leads to low
246   photolysis rates (Segura et al 2005). Very high concentrations of $CH_4$ and $N_2O$
247   could, in principle, push the outer edge of the habitable zone past the $CO_2$
248   condensation limit, in which case the spectral dependence of the frozen phases of
249   $CO_2$, $N_2O$, and even $CH_4$ might be important for determining the actual outer

250     edge of the habitable zone around M-stars. We suggest that this could be a source

251     for future research.

252

253     **Conclusions**

254     We have shown that considering the large reduction in water ice and snow albedo

255     at wavelengths longer than 1 μm significantly lowers the mean albedos of ice and

256     snow on planetary surfaces orbiting M-stars. The effect is because M-stars emit a

257     significant fraction of radiation at these longer wavelengths (see Figure 1). The

258     effect of such spectral dependence can move the habitable zones of planets

259     orbiting M-stars outward by 10-30% in terms of distance from the star and

260     increase the chance of finding habitable planets orbiting M-stars.

261

262     More detailed calculations of the effect will have to take into account spatially

263     varying clouds and emissivity (because of water vapor, especially if there are

264     large thermal contrasts on the planet) and employ three-dimensional climate

265     models. However, that does not change the conclusion that any terrestrial planets

266     that orbit the large majority of main sequence stars will not have bright ice or

267     snow caps, as is the case on Earth, and as such will not exhibit any significant

268     snow/ice albedo climate feedback.

269

270

**Acknowledgements**

271

272 MJ is supported by the U.K. National Centres of Atmospheric Science (NCAS)

273 Climate. The authors would like to thank Jonathan Fortney for the M-star spectral

274 data as well as Laurance Doyle for his helpful comments. We acknowledge the

275 extremely useful comments of Stephen Warren and the other reviewers of the

276 submitted manuscript.

277

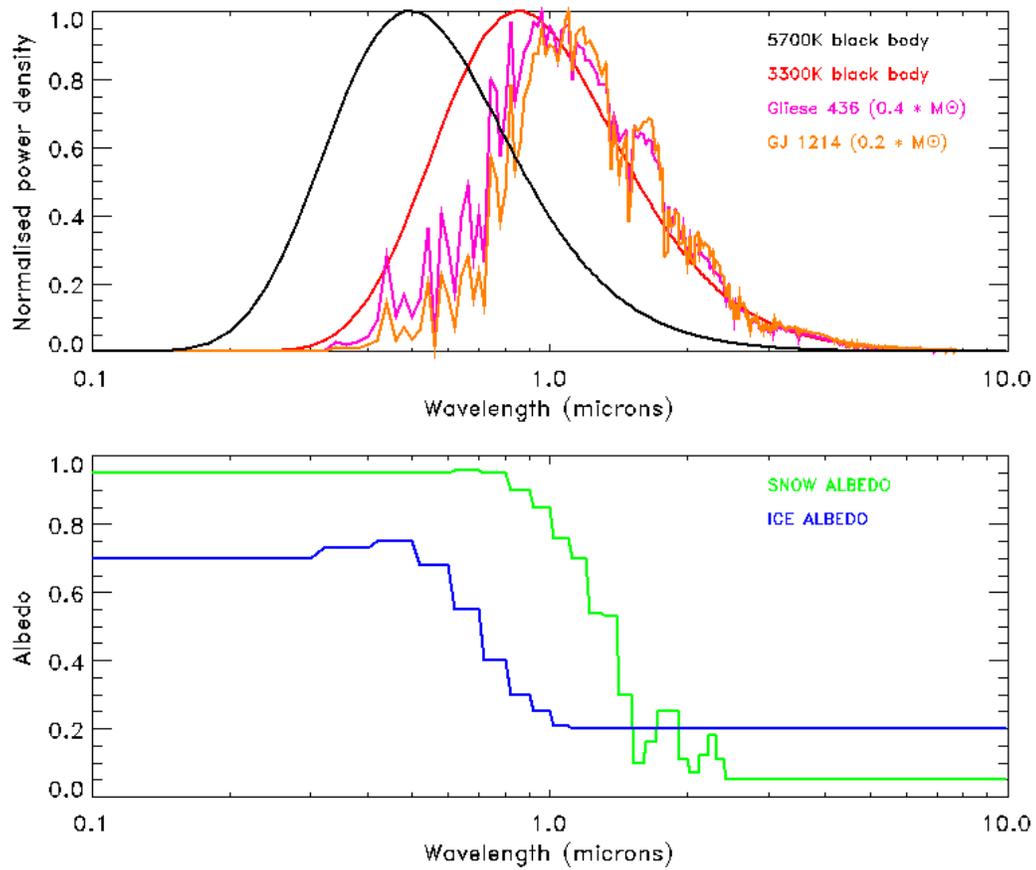

343

344    **FIGURE 1**

345    Top panel: the spectral distribution of energy for differing emitters of radiation,

346        normalized by the peak value. The curves correspond to: black- 5700K black

347        body, similar to the Sun; red- 3300K black body, which is an idealized

348    representation of an M-dwarf that is approximately 40% as massive as the Sun;

349    purple- Gliese 436, which is approximately 40% as massive as the Sun; orange-

350    GJ 1214, which is approximately 20% as massive as the Sun. The non-Planckian

351        behavior of the M-star spectra is because of absorption by species such as

352      Titanium oxide (TiO) at visible wavelengths, and windows where upwelling

353      radiation is coming from deeper in the stellar interior where temperatures are

354      higher at near-infrared wavelengths.

355      Bottom panel: the spectral distribution of snow (green) and ice (blue) bond

356      albedos employed in the model.

357

358

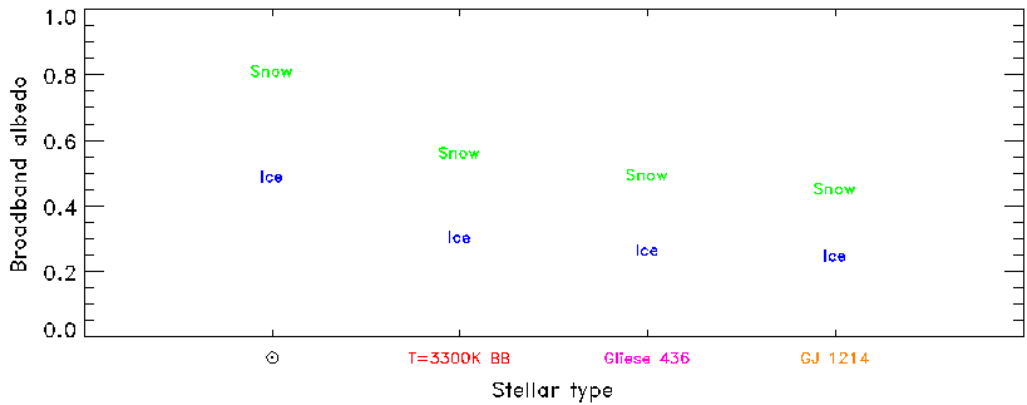

359

360      **FIGURE 2**

361      The snow and ice broadband albedos for the 4 stellar types shown in Figure 1 (top

362      panel), assuming 100% snow or ice cover. Broadband albedos are calculated by

363      weighting the spectrally varying snow and ice albedos in Figure 1 (bottom panel)

364      with use of the normalized spectral energy distributions of Figure 1 (top panel).

365      The colors of the spectral types are consistent with Figure 1 (top panel).

366

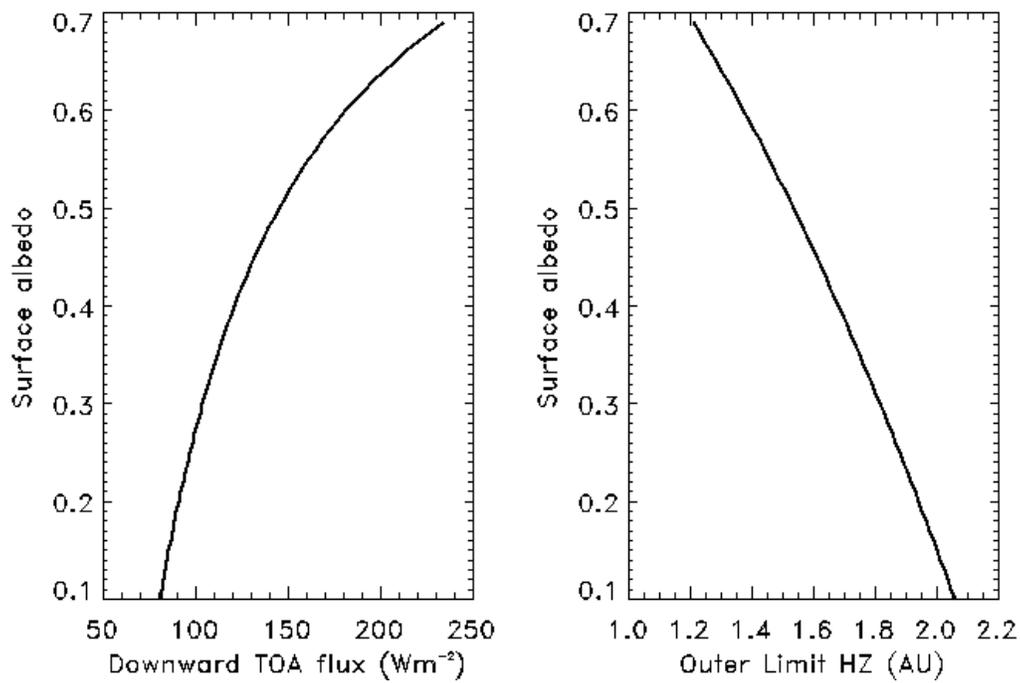

367

368        **FIGURE 3**

369        Left panel: Downward TOA stellar flux at which a grey atmosphere with an

370        emissivity ε of 0.8 has a surface temperature of 200K vs surface albedo assuming

371        no atmospheric absorption. Right panel: As left panel but x-axis is the distance

372        from the parent star in AU assuming a stellar flux of 342 Wm$^{-2}$ at 1 AU and

373        stellar flux is proportional to (orbital distance)$^{-2}$

374